# *Stable* Co-Catalyst-Free Photocatalytic H$_2$ Evolution From Oxidized Titanium Nitride Nanopowders

Xuemei Zhou,[a] Eva M. Zolnhofer,[b] Nhat Truong Nguyen,[a]  Ning Liu,[a]

Karsten Meyer,[b]  Patrik Schmuki*[a]

---

[a]  Xuemei Zhou, Nhat Truong Nguyen, Ning Liu, Patrik Schmuki

Department of Materials Science WW-4, LKO, University of Erlangen-Nuremberg, Martensstrasse 7, 91058 Erlangen, Germany;

E-mail: schmuki@ww.uni-erlangen.de

[b]  Eva M. Zolnhofer, Karsten Meyer

Friedrich-Alexander University Erlangen - Nürnberg (FAU), Department of Chemistry and Pharmacy, Inorganic Chemistry, Egerlandstr. 1, 91058 Erlangen, Germany.

**Abstract:**

In the present work, a simple strategy is used to thermally oxidize TiN nanopowder (~20 nm) to an anatase phase of a $TiO_2:Ti^{3+}:N$ compound. In contrast to the rutile phase of such a compound this photocatalyst provides photocatalytic activity for hydrogen evolution under AM1.5 conditions – this without the use of any noble metal co-catalyst. Moreover the photocatalyst is active and stable over extended periods of time (tested for 4 months). Importantly, to achieve successful conversion to the active anatase polymorph, sufficiently small starting particles of TiN are needed. The key factor for catalysis is the stabilization of the co-catalytically active $Ti^{3+}$ species against oxidation by nitrogen present in the starting material.

In 1972, Fujishima and Honda have reported groundbreaking work on the photolytic cleavage of water to $H_2$ and $O_2$. [1] Since then, the concept of using solar light and a suitable semiconductor to generate the fuel of the future, $H_2$, has received tremendous scientific attention. In their experiments, Fujishima and Honda used a two electrode approach with a $TiO_2$ single crystal as an illuminated photoanode and a Pt sheet as a counter electrode for hydrogen evolution (i.e., a photoelectrochemical configuration).

More straightforward than using a two-electrode approach is to use a suspension of nanoparticles (that is without external electrochemical bias), where the photogenerated holes and electrons from the same particle react with the surrounding water. However, under these conditions, the presence of a suitable co-catalyst on the $TiO_2$ particles is required in order to efficiently generate $H_2$. Over the past decades in virtually any investigation on photocatalytic hydrogen evolution, noble metals, such as Pt, Au, Pd[2], have been used as they present most effective co-catalysts due to their abilities to act as electron transfer mediator and recombination center for $H_2$. Efforts to replace these expensive noble metal co-catalysts are limited to only a few reports [2a, 3].

A key factor to form noble-metal-free active material may be the formation of a specific configuration of $Ti^{3+}/O_v$ ($O_v$=oxygen vacancies) on anatase particles. However, well established reduction treatments of $TiO_2$ known to form $Ti^{3+}/O_v$ species (such as heat treatment in vacuum, Ar, $Ar/H_2$, ion bombardment or cathodic reduction) [2a, 4] lead to a non-stable co-catalytic effect due to the easy oxidizability of $Ti^{3+}$ in air or aqueous environment. [3a-b, 5]

An alternative approach to form $Ti^{3+}$-rich $TiO_2$ is, in principle, to perform a controlled partial oxidation of Ti(III) compounds instead of a partial reduction of $TiO_2$. Nevertheless, from previous work [5, 6] we know that the two key requirements to obtain a stable and active catalyst are: *i)* the stabilization of the remaining Ti(III) against further oxidation, and *ii)* the formation of an *anatase* type of polymorph $TiO_2$ by oxidation (rutile is reported not to be active). In order to address requirement *i)*, one may consider that theoretical work combined with experimental evidence indicate that nitrogen species situated in interstitial and/or substitutional positions are capable of stabilizing $Ti^{3+}$ centers by charge transfer resonance - as for example suggested by Livraghi and Hoang et al. [7]. Therefore the question arises, if an optimized oxidation of TiN may lead to a nitrogen stabilized $Ti^{3+}$ configuration and additionally can result, when oxidized, in an anatase form of $TiO_2$ (and thus also addresses *ii)*).

Some previous attempts to thermally oxidize TiN were mainly undertaken to create "N-doped" $TiO_2$ with the intent to alter the light absorption properties towards a visible light response, i.e., to provide an alternative to other nitrogen doping methods (such as ammonolysis [8], ion implantation [9] and co-precipitation [10]) with the aim to create band-gap narrowing in $TiO_2$ (ascribed to a mixing of N 2p - states with the O 2p valence band states of $TiO_2$ [11a]). Previous work on the oxidation of TiN used either comparably large particles or surface films that in any reported case evolved, upon thermal oxidative treatment, to a rutile phase [11]. This polymorph is undesired for our purposes, since in previous works a $H_2$ evolving co-catalytic $Ti^{3+}$-state was only obtained for anatase. [6a, 12a]) Rutile is the thermodynamically stable form of titania under a broad range of conditions. However, various theoretical work reports that at the nanoscale (typically < 30 nm), anatase can become the thermodynamically stable polymorph [3a, 12].

In the present work, we thus explore the use of TiN nanoparticles (≈20 nm) for a controlled partial thermal oxidation and investigate the photocatalytic properties, composition and structure at various stages of oxidation. We show that indeed such TiN nanoparticles can be oxidized to an anatase-type titania photocatalyst that carries stable, intrinsic co-catalytic centers for hydrogen evolution, facilitating the stable photocatalytic $H_2$ production at enhanced rates without the use of nobel metal co-catalysts. This finding is illustrated in Fig. 1. Fig. 1a shows the measured initial photocatalytic hydrogen evolution rates under AM1.5 (100 mW/cm$^2$) conditions for various oxidized TiN powder samples as well as for various reference samples (namely anatase and reduced anatase). Experimental details are given in the SI (see also Fig. S1). In order to produce oxidized TiN, commercial TiN (20 mg) of an average grain size of ≈20 nm was annealed in air under different conditions (300 °C – 450 °C for 30 min – 9 h). A suspension of this sample was prepared and the photocatalytic activity for $H_2$ evolution was measured over time. Partially reduced anatase samples (fabricated by thermal Argon or Argon/$H_2$ treatment s[3b, 5e-f, 13]) were used as reference samples in order to provide a comparison to conventionally reduced $Ti^{3+}$-containing material without any nitrogen-stabilizing effects. As shown in Fig.1a, various annealing treatments of TiN in air lead to a product that yields a significantly enhanced photocatalytic activity for $H_2$ evolution compared to pure anatase (i.e. without using any noble metal co-catalyst). While for pure titanium nitride, or samples that were annealed up to 300 °C, no photocatalytic hydrogen could be detected, treatments with temperatures higher than 350 °C, clearly showed photocatalytic activity for $H_2$ evolution. By investigation of a range of annealing conditions - varying temperature and time (see Fig. S1, ESI) - we have found that the optimal conditions for air annealing are either at 400 °C for 1 h, or similarly at 350 °C for 5 h.

Most importantly, Fig.1b shows that this oxidized TiN material provides an excellent stability of the intrinsic co-catalytic effect over time and in repeated experiments. In contrast, reduced anatase samples (Ar, Ar/$H_2$) show some initial activity (Fig. 1b) that, however, is quickly lost after 2 or 3 photocatalysis experiments. In fact, for TiN the data show that the $H_2$ evolution rate remains constant over a period of over 4 months (longest measured duration). These findings suggest that the combination of nitrogen and $Ti^{3+}$ states leads to a remarkably stable catalytic center for $H_2$ evolution without the use of a co-catalyst. Fig. 1c shows the HRTEM image and the corresponding SAED pattern (Fig. 1d) from such an activated grain of the photocatalyst. The SAED pattern confirms the presence of a mixture of anatase and TiN after annealing. In the HRTEM in Fig. 1c, lattice fringes of d = 0.34 nm are determined, which correspond to a typical lattice spacing of anatase (101). Moreover, the thermal treatment does not change the morphology nor the size of the particle significantly (Fig. S2). It may also be noted that the conversion of TiN to the active catalyst material can easily be followed by eye, as the initially black color of the TiN is changed to grey by annealing (Fig. S2g).

In order to follow the structural and compositional changes during the annealing treatments, the different oxidation stages were characterized by XRD (Fig. 2 and Fig. S3), Raman spectroscopy (Fig. S4) and XPS (Fig. 3 and Fig. S5).

Fig. 2 shows the XRD patterns of the nanoscale TiN samples (SEM in Fig. S2a) annealed in air in the temperature range from 300 °C to 450 °C for 1 h. The XRD pattern of the originally 20 nm titanium nitride materials can be assigned to osbornite, cubic phase (PDF card No. 00-038-1420), presenting a main peak at 42.6 °. With increasing temperature, the intensity of the XRD peak for TiN decreases while the peak intensity of the $TiO_2$ anatase phase (PDF card No. 00-021-1272, Tetragonal) increases. For annealing at 400 °C for 1 h, clear anatase peaks become apparent; and at a temperature of 450 °C, also traces of rutile (PDF card No. 00-021-1276, Tetragonal) become visible, while no TiN peaks can be detected anymore. The formation of anatase in the TiN particles can also be observed by Raman spectroscopy (Fig. S4), with characteristic absorption bands at ≈ 145 $cm^{-1}$, 396.0 $cm^{-1}$, 514.6 $cm^{-1}$ and 637.7 $cm^{-1}$ (Fig. S4a). For the nanoscale-catalyst-material, peak shifts and -broadening (Fig. S4b) (TiN 400 °C – 1 h) compared to pure anatase reference powder is observed. This may be related to confined lattice vibrations and structural defects [14] for the nitrogen doped $TiO_2$ nanomaterial. In order to study annealing-time related effects of the conversion, we kept a TiN (20 nm) sample at 350 °C for various times (Fig. S3b). With increasing annealing time, the anatase content ($X_a$) can be increased from 0.043 to 0.96. This is also observed at an annealing temperature of 400 °C, when annealing times from 30 min to 5 h are used, the content of anatase increases from 0.20 to 1.0.

Fig. 3 provides high resolution XPS spectra of selected samples before thermal treatment and at different oxidation states, clear alterations of Ti2p peaks as well as of N1s peaks are observed after annealing in air. Fig. 3 shows the Ti2p peak for TiN with a strong typical $Ti^{3+}$ tail below 457 eV. During annealing at 350 °C, the intensity of this tail decreases and for annealing for longer than 3h, the $Ti^{3+}$ is almost fully oxidized to $Ti^{4+}$ (within the penetration depth of XPS). Correspondingly, significant changes of the N1s peak can be observed before and after annealing as shown in Fig. S5. The high resolution N1s peak for TiN located at binding energies of 396.9 eV and 395.8 eV, is commonly ascribed to a nitride compound and to substitutional nitrogen in $TiO_2$. After annealing at 350 °C, the intensity of peaks between 396 eV – 397 eV decreases and new peaks at a binding energy of ≈ 402 eV appear. These peaks can be assigned to surface bound $NO_x$-species that probably are released from the lattice during oxidation [15]. After annealing at 350 °C for 5 h or higher temperatures, the only traceable N peak is this 402 eV signature. However, in XRD, for the TiN samples annealed at 400 °C for 1 h and TiN 350 °C for 5 h, the crystalline TiN diffraction peak at 42.6 ° is still visible after the thermal treatment. This indicates that the majority of TiN remains as a core of the particle while at the stronger oxidized surface, nitrogen is only left at very low concentrations, [i.e. below the detection limit of XPS (≈ 1 at. %)]. During the thermal annealing, a change in the high resolution O1s peak (Fig. S5a) can also be seen. Generally, the intensity of the O1s peak located at a binding energy of 529.8 eV (oxide) increases, which can be ascribed to the formation of $TiO_2$, while the peak intensity at a higher binding energy (usually assigned to surface OH) decreases. Overall, the XPS data are well in line with the formation of $TiO_2$ surrounding a TiN core. A quantitative evaluation of peak intensities is compiled in Table S1. It is clear that with an increase in annealing time and temperature, the ratio of N/O in the TiN samples decreases from 0.456 to 0.0133, in agreement with a transition from TiN to $TiO_2$.

Additionally, we characterized the samples with electron paramagnetic resonance spectroscopy (EPR) that is widely used in the characterization of paramagnetic centers, such as $Ti^{3+}$ or oxygen vacancy species present in $TiO_2$ [16]. Fig. 4a shows the EPR spectra for the active TiN-based catalyst generated by an air-heat treatment (400 °C – 1 h) and a reference sample, where $Ti^{3+}$-species were generated by an Ar-heat treatment (500 °C – 3 h). The EPR spectra were measured directly after the thermal treatment and after ageing the samples for four months (air exposure and regular $H_2$ photo-activity measurements). The EPR spectra of all four samples exhibit a signal at a g-value of g = 1.997, which is attributed to a $Ti^{3+}$ defect in the sample, as reported in the literature [16e, 17]. In both fresh samples (TiN400-1h fresh and Ar-anatase fresh) as well as the aged sample (TiN400-1h after 4 months), additional low-intense signals are shown in the range of g = 1.97 – 2.03 (obtained at RT under dark conditions). Upon illumination at 100 K, the intensity of the additional signals is highly increased due to the photochemical generation of additional paramagnetic centers (figure 4b). The better signal-to-noise ratio facilitates the simulation of the EPR signals, which reveals the presence of two species (figure 4c): one highly isotropic signal with g-values of g1 = 2.0031, g2 = 2.0026 and g3 = 2.0012 (species A, assignable to "F-centers" or oxygen vacancies [17]) and one rhombic signal with g-values of g1 = 2.0220, g2 = 2.0056, g3 = 1.9810 (species B, assignable to a $Ti^{3+}/N_b$• characteristic feature [7, 18] - see also SI.). Most importantly, after ageing, a clear difference in the spectra between the TiN400°C-1h sample and the Ar-anatase sample is evident: while the signals for the nitride sample remains stable upon ageing for four months, the reference sample shows a strong decrease of the signal intensity over 20 days. This indicates that re-oxidation of the locally reduced anatase sample eliminates the $Ti^{3+}/O_v$ centers to a large extent; i.e., below the resolution limit of EPR. The high stability of the TiN 400°C-1 h sample is also evident when investigated at 100 K under illumination in an air atmosphere (Fig. 4) and other conditions (see SI, Fig. S6): in all measurements, the fresh TiN 400°C-1 h (Fig.4b) as well as the four-months-aged sample (Fig.4c) give virtually identical signals of comparable intensity and hence are in line with the long-term stability of the catalytically active species (i.e. the hydrogen evolution activity in Fig.1b). The EPR measurements also strongly support concepts of charge-transfer resonance [7b] being the origin of the N-stabilization effect of $Ti^{3+}$ in thermally oxidized TiN nanoparticles.

In summary, the present work shows that TiN nanoparticles (if sufficiently small) not only can be successfully converted to an anatase based $TiO_2$ photocatalyst, but more importantly, the material shows a highly stable photocatalytic $H_2$-production (tested over 4 months) performance without the use of any noble metal co-catalyst. EPR spectra further confirm the remarkable stability of photocatalytic activity observed for these samples over time and further support the concept of a $Ti^{3+}$ related defect structure capable to act as a co-catalytic center for $H_2$ evolution.

**Acknowledgements**

The authors would like to thank Manuel Schweiger and Prof. Dr. Jana Zaumseil for the Raman measurements. We would also like to thank ERC, DFG and the EAM cluster of excellence for financial support.

**Keywords:** Anatase • $H_2$ evolution • nitrogen stabilization • $Ti^{3+}$ • nanopowders

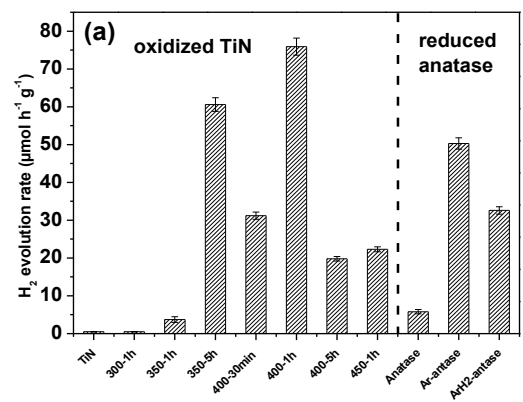

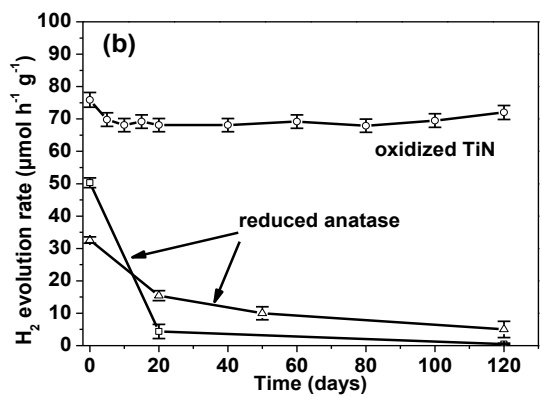

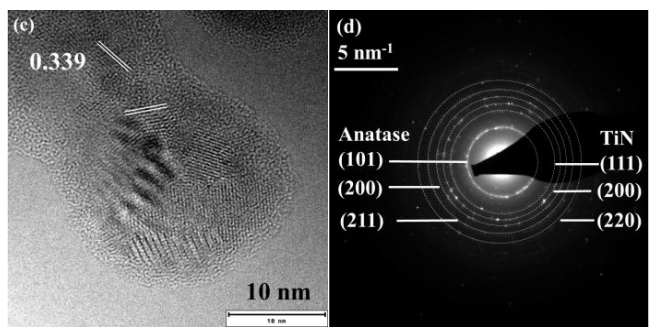

**Figure 1.** (a) Open circuit hydrogen generation for oxidized TiN nanopowders treated at different temperatures in air and comparison to reduced anatase samples as a reference. (b) Stability of photocatalytic $H_2$ evolution for oxidized TiN (400 °C, 1 h) and reduced anatase (Δ: Ar/$H_2$ 500 °C, 3 h □: Ar 500 °C, 3h). $H_2$ evolution experiments were performed under AM1.5 (100mW/cm2) at room temperature in 50% methanol/$H_2O$ electrolyte without the presence of any co-catalyst. (c) TEM and (d) corresponding SAED patterns for oxidized TiN (TiN 350 °C – 3 h, i.e. optimized photocatalyst).

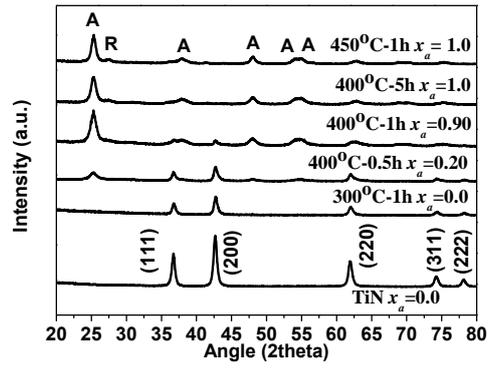

**Figure. 2**. XRD patterns for untreated titanium nitride and after annealing at 300 °C, 400 °C, 450 °C. *Xa* is the fraction of anatase in the mixture by calculation from the integrated intensities of the (101) reflection of anatase and the (200) reflection of titanium nitride.

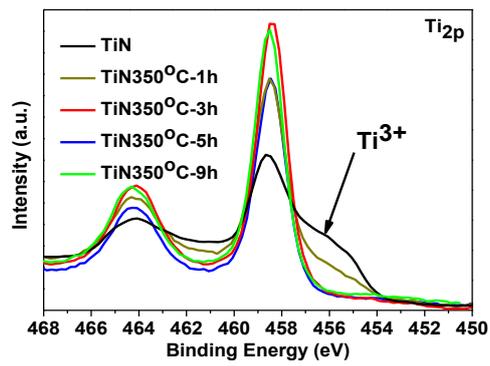

**Figure. 3**. High resolution XPS Ti2p peaks for TiN nanopowders and oxidized TiN at 350°C for 1h to 9h, respectively.

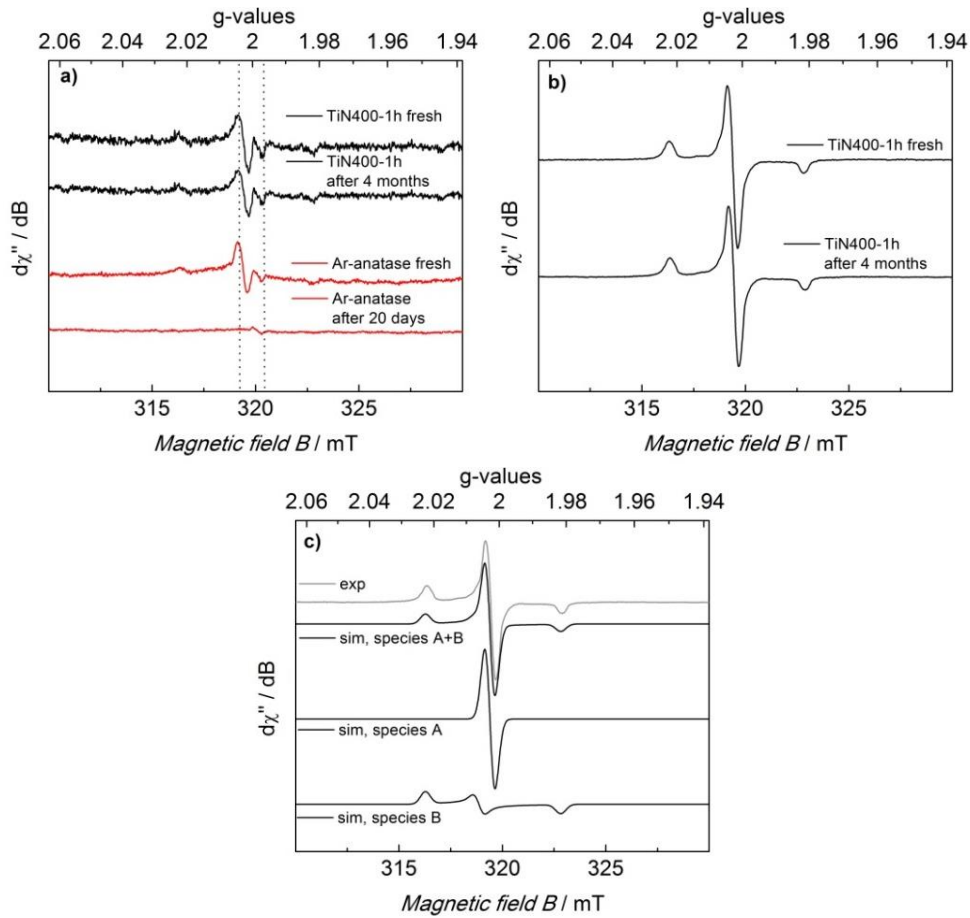

**Figure. 4.** CW X-band EPR spectra of oxidized titanium nitride (400 °C, 1h) directly after preparation and after ageing for 4 months, recorded in the solid state at RT and at 100 K. Experimental conditions: microwave frequency = 8.96 GHz, modulation width = 0.5 mT, microwave power = 1 mW, modulation frequency = 100 kHz, time constant = 0.1 s. (a) Comparison of oxidized titanium nitride to Ar-reduced anatase (freshly prepared and aged) measured at RT. (b) EPR spectra of TiN400°C -1h, freshly prepared and after ageing for 4 months, measured at 100 K under illumination. (c) EPR spectrum of TiN400°C taken at 100 K under illumination (grey line) and its simulation (two species, black lines). Simulation parameters: Species A: weight = 1.00, S = 1/2, g1 = 2.0031, g2 = 2.0026, g3 = 2.0012, W1 = 0.31mT, W2 = 0.25mT, W3 = 0.28mT. Species B: weight = 2.75, S = 1/2, g1 = 2.0220, g2 = 2.0056, g3 = 1.9810, W1 = 0.28mT, W2 = 0.27mT, W3 = 0.30mT.

# Supporting Information

# *Stable* Co-Catalyst-Free Photocatalytic H$_2$ Evolution From Oxidized Titanium Nitride Nanopowders

Xuemei Zhou[1], Eva M. Zolnhofer[2], Nhat Truong Nguyen[1], Ning Liu[1], Karsten Meyer[2], Patrik Schmuki[1]*

[1]Department of Materials Science WW-4, LKO, University of Erlangen-Nuremberg, Martensstrasse 7, 91058 Erlangen, Germany;

[2]Friedrich-Alexander University Erlangen - Nürnberg (FAU)

Department of Chemistry and Pharmacy, Inorganic Chemistry

Egerlandstr. 1, 91058 Erlangen, Germany

*Corresponding author. Tel.: +49 91318517575, fax: +49 9131 852 7582

Email: schmuki@ww.uni-erlangen.de

**Experimental Section**

Titanium nitride (TiN, 20 nm, 99.0 %, Goodfellow) was used as a precursor without any further treatment. For oxidation, generally 20 mg of the nanopowder was placed into a ceramic boat and annealed in a muffle tube furnace (Heraeus R07/50, 220 V, 13.6 A, 50/60 Hz, 3.0 kW) under air atmosphere at various temperatures (300 °C – 450 °C) for various times (30 min – 9 h). After annealing, the powders show variations of grey color. For reference samples, we reduced anatase $TiO_2$ (<25nm, Sigma Aldrich) by thermal annealing at 500 °C for 3 h in the tube furnace under an argon or $Ar/H_2$ atmosphere at a flow rate of 6 L/h.

**Characterization**

XRD spectra were collected using an X'pert Philips PMD diffractometer with a Panalytical X'celerator detector, using graphite-monochromatized CuKa radiation ($\lambda$=1.54056Å). The $TiO_2$ (anatase) ratio was determined by a technique described in Ref. [S1] using the following equation:

$$x_a = \frac{1}{1 + 1.21\,(I_o/I_a)}$$

where Xa is the fraction of anatase in the mixture, while $I_a$ and $I_o$ are the integrated intensities of the (101) reflection of anatase and the (200) reflection of titanium nitride osbornite. Factor 1.21 was obtained by calculation of the integrated area versus ratio from mixture of pure TiN and anatase powders.

Raman spectra were acquired using a Renishaw in Via Reflex Confocal Raman Microscope with an excitation laser wavelength of 532 nm (High-resolution).

An X-ray photoelectron spectrometer (XPS, PHI 5600 XPS spectrometer, USA) was used for detection of elements and chemical states in the compound. XPS spectra were acquired using Al standard X-rays with a pass energy of 23.5 eV. All XPS element peaks were shifted to a C1s standard position (284.8 eV).

EPR spectra were recorded on a JEOL continuous wave spectrometer JES-FA200 equipped with an X-band Gunn oscillator bridge, a cylindrical mode cavity, and a helium cryostat. The samples were measured in the solid state in an air atmosphere (or argon) in air-tight J. Young quartz EPR tubes with similar loading. Background spectra were obtained on empty tubes at the same measurement conditions. The spectra shown were measured with the following parameters: Temperature RT or 100 K, microwave frequency 8.960 GHz (1) and 8.966 GHz (2), modulation width 0.5 mT, microwave power 1 mW, modulation frequency 100 kHz, and time constant 0.1 s. Spectral simulation was performed using the program W95EPR written by F. Neese [S2]. A LOT 150W Xe-OF arc lamp (wavelength range 200-900 nm) was used for the illumination experiments.

Photocatalytic hydrogen generation was measured under open circuit conditions from an aqueous methanol solution (50 vol%) under AM 1.5 (100 mW/cm$^2$) solar simulator illumination. The amount of $H_2$ produced was measured using gas chromatography (GC). $H_2$ detection was carried out in the gas phase over the illuminated liquid in a closed reactor on a Varian gas chromatograph (SCHIMADZU GAS CHROMATOGRAPH GC-2010 plus) with a TCD detector. To prepare suspensions for $H_2$ measurements, 2 mg sample powders were dispersed in 10 mL of DI water/methanol (50/50 v%) with ultra-sonication for 15 min. During illumination, the suspensions were continuously stirred. For rate determination, data were taken at 3 h, 8 h, 12 h and 20 h during solar simulator irradiation. The stability in an air atmosphere was tested with samples after several days of air exposure. Samples were exposed to air and their photocatalytic activity was checked after different intervals of time.

**Additional material:**

Fig. S1a shows that with increasing annealing time at 350 °C from 1 h to 9 h one can reach various oxidation stages of TiN. For 5 h annealing, a photocatalytic activity of 60 µmol/(g·h) is reached, very close to the optimized condition at 400 °C for 1 h (See Fig.1 main text). Fig. S1b shows additionally that the $H_2$ production increases linearly with time.

Fig. S2a gives an overview SEM image of TiN nanopowders, as purchased, i.e. without any post-treatment. These particles have an average particle diameter of approx. 20 nm. A HRTEM image of such TiN nanoparticles − (Fig. S2e) presents a typical titanium nitride particle − shows lattice fringes corresponding to crystal faces of (200) and (111) with a spacing of d = 0.211 nm and d = 0.245 nm, respectively. This is also shown in Fig. S2f (SEAD), the distance and strength of respective patterns are in line with literature [S3] as well as the XRD results (Fig.2a). SEM images for reference commercial anatase powders are shown as delivered (Fig.S2c) and after annealing in Ar (Fig.S2d) – the results illustrate that the used particle size is comparable with TiN. The XRD patterns for Ar-anatase and Ar/$H_2$-anatase are presented in Fig.S3a. In both cases after treatments at 500 °C, mainly anatase, and only a small amount of rutile is formed in Argon.

EPR spectra for different oxidation stages were taken at RT and at 100 K with and without illumination in an air and in an argon atmosphere. They clearly show that the thermal treatment of titanium nitride has led to changes in the material. The EPR spectra of the TiN taken before and after illumination are silent. Fig.S6a shows the EPR spectra of a selection of samples before illumination at RT. With the increase of the annealing duration i.e. formation of anatase in the titanium nitride, the signature at g ≈ 2 becomes more obvious within the limits of the low signal intensity at this temperature. In the EPR spectra recorded at a temperature of 100 K under illumination, the changes in the spectra become clear due to a higher signal intensity and better signal-to-noise ratio: with then increase of annealing time, a higher amount of oxidation, which suggests a higher content of paramagnetic centers, is displayed. The spectra for pure anatase were also recorded at RT (Fig.S6a), which presents main signals at g ≈ 2.

For oxidized TiN samples, under illumination, photo generated electrons react with $TiO_2$ [S4-5]. In the band gap of anatase, localized paramagnetic $N_b^\bullet$ and diamagnetic $N_b^-$ are generated [S6]. Energetically they lie near (ΔE <1eV) the valence band. Under illumination, they react as following [S6]:

$$N_b^\bullet + O_2(g) \xrightarrow{h\nu} N_b^+ + O_2^{\bullet-} (surface) \ in \ air$$

$$N_b^- + O_2(g) \xrightarrow{h\nu} N_b^\bullet + O_2^{\bullet-} (surface) \ in \ air$$

$$N_b^- \xrightarrow{h\nu} N_b^\bullet + e^- \ in \ argon$$

$$N_b^\bullet \xrightarrow{h\nu} N_b^+ + e^- \ in \ argon$$

The signal located at g1=2.0220 is assigned to $O_2^{\bullet-}$ radicals on the particle surface (Fig.4d and Fig. S6c) [S6-7]. This assignment is more clear when spectra are recorded at 10K (liquid He) (Fig.S6d). In argon atmosphere (Fig.S6c), the signal can be ascribed to surface $Nb^{\bullet}$. The results thus are in line with literature and suggest the nitrogen located in the band gap of anatase can stabilize crystal defects such as $Ti^{3+}$ according to charge resonance stabilization:

$$N_b^{\bullet} + Ti^{3+} \leftrightarrow N_b^{-} + Ti^{4+}$$

Table S1. Quantitative atomic ratios calculated from XPS peaks of different elements.

| Samples | Ti | N | O | N/O | N/Ti |
|---|---|---|---|---|---|
| TiN | 21.34 | 16.69 | 36.59 | 0.456 | 0.782 |
| TiN300°C-1h | 25.64 | 13.24 | 45.20 | 0.293 | 0.516 |
| TiN350°C-1h | 25.06 | 7.06 | 50.80 | 0.139 | 0.282 |
| TiN350°C-3h | 24.77 | 2.11 | 56.88 | 0.0371 | 0.0852 |
| TiN350°C-5h | 23.45 | 1.25 | 54.56 | 0.0229 | 0.0533 |
| TiN350°C-9h | 23.71 | 0.73 | 56.89 | 0.0128 | 0.0308 |
| TiN400°C-1h | 23.25 | 0.75 | 56.19 | 0.0133 | 0.0323 |

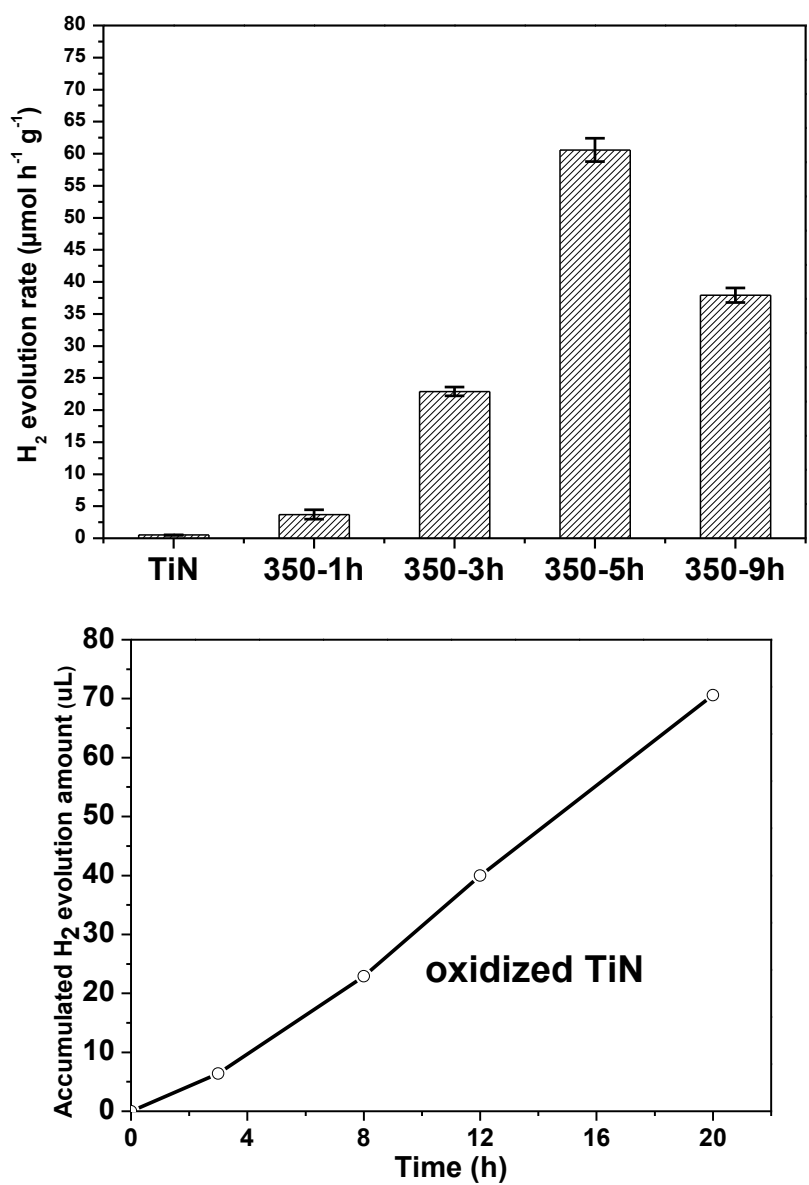

Fig. S1. Open circuit hydrogen generation for oxidized TiN nanopowders treated at 350°C for various times. (b) Open circuit hydrogen generation for oxidized TiN (400°C, 1h) as a function of irradiation time. $H_2$ evolution experiments were performed under AM1.5 (100mW/cm$^2$) at room temperature in 50% methanol/$H_2O$ electrolyte without the presence of any co-catalyst.

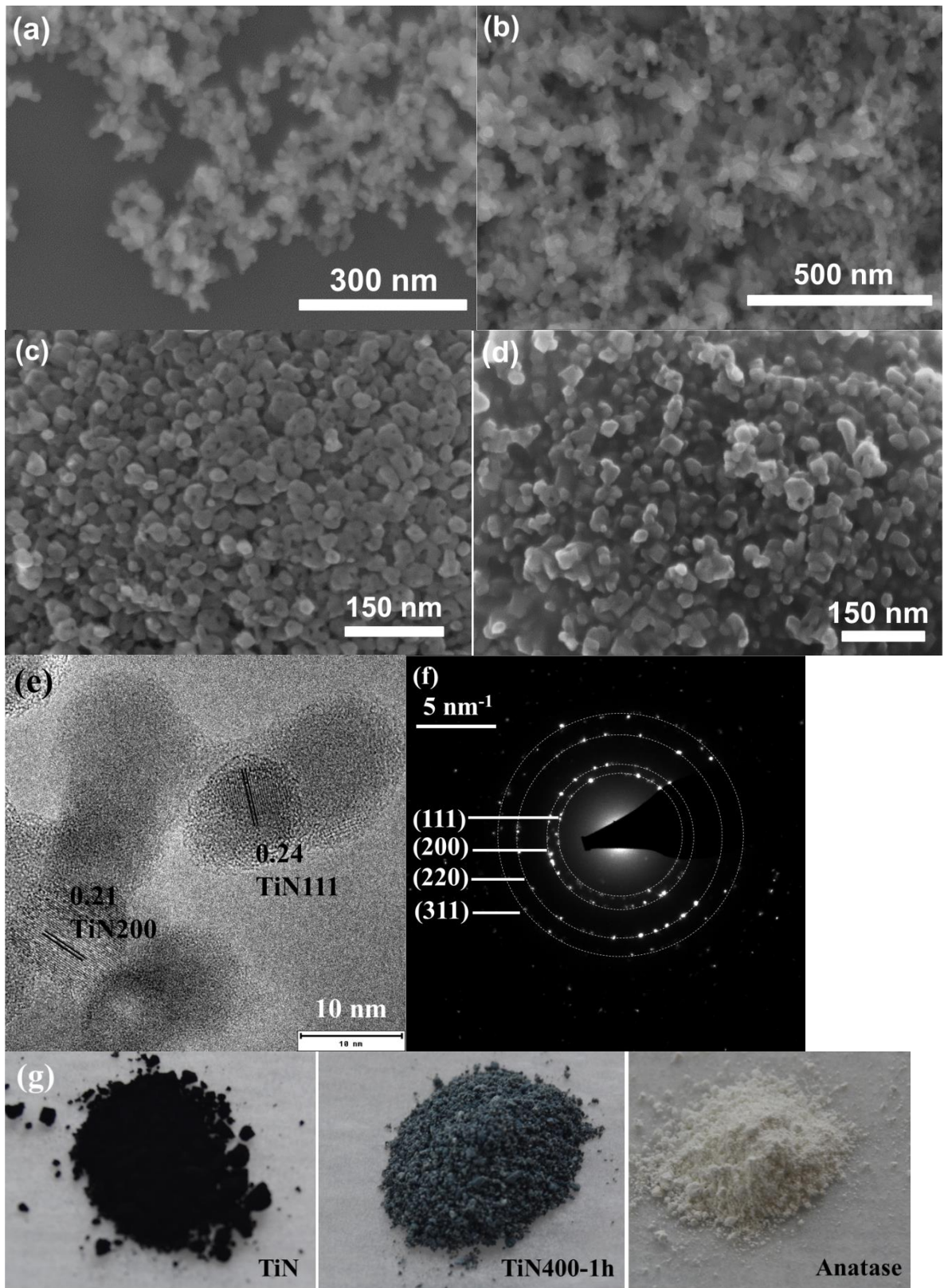

Fig.S2. SEM images of (a) TiN, (b) oxidized TiN at 400°C for 1h, (c) anatase nanopowders and (d) reduced anatase in Ar at 500°C for 3h, (e) HR-TEM and (f) SEAD patterns for TiN nanopowders. (g) Optical images of TiN, oxidized TiN at 400°C for 1h and anatase.

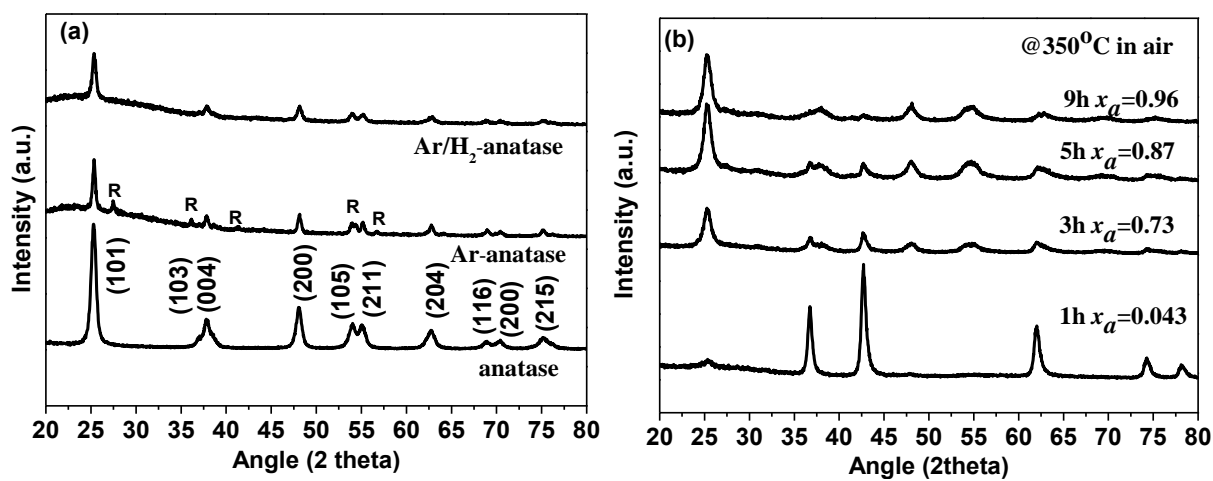

Fig.S3. (a) XRD for commercial anatase, anatase reduced in argon or argon/$H_2$ at 500°C for 3h, respectively. (b) XRD patterns for titanium nitride after annealing at 350°C for 1h, 3h, 5h and 9h. *Xa* is the fraction of anatase in the mixture by calculation from the integrated intensities of the (101) reflection of anatase and the (200) reflection of titanium nitride.
22

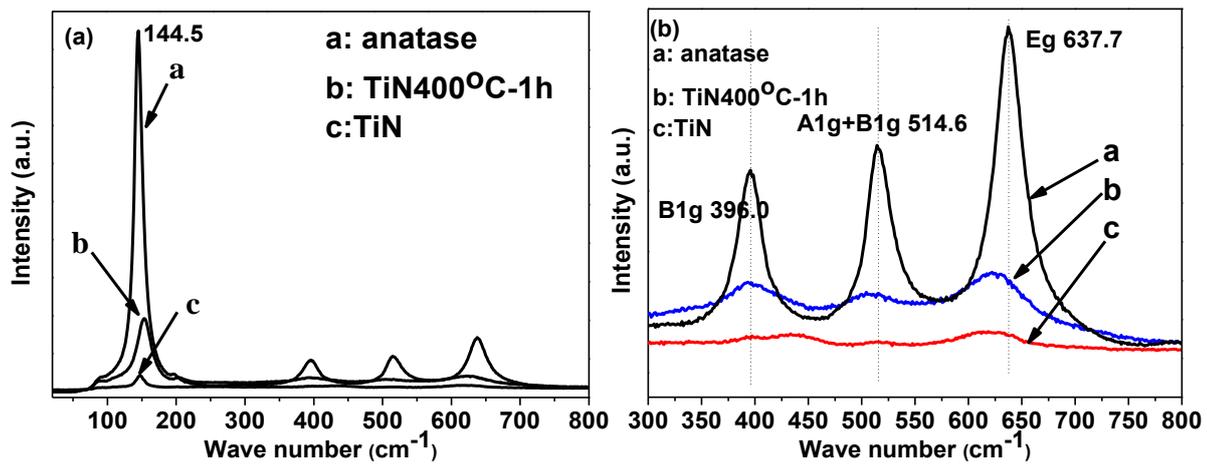

Fig.S4. Raman spectra in overview a) and (b) high resolution of commercial anatase, TiN and oxidized TiN(400°C, 1h).

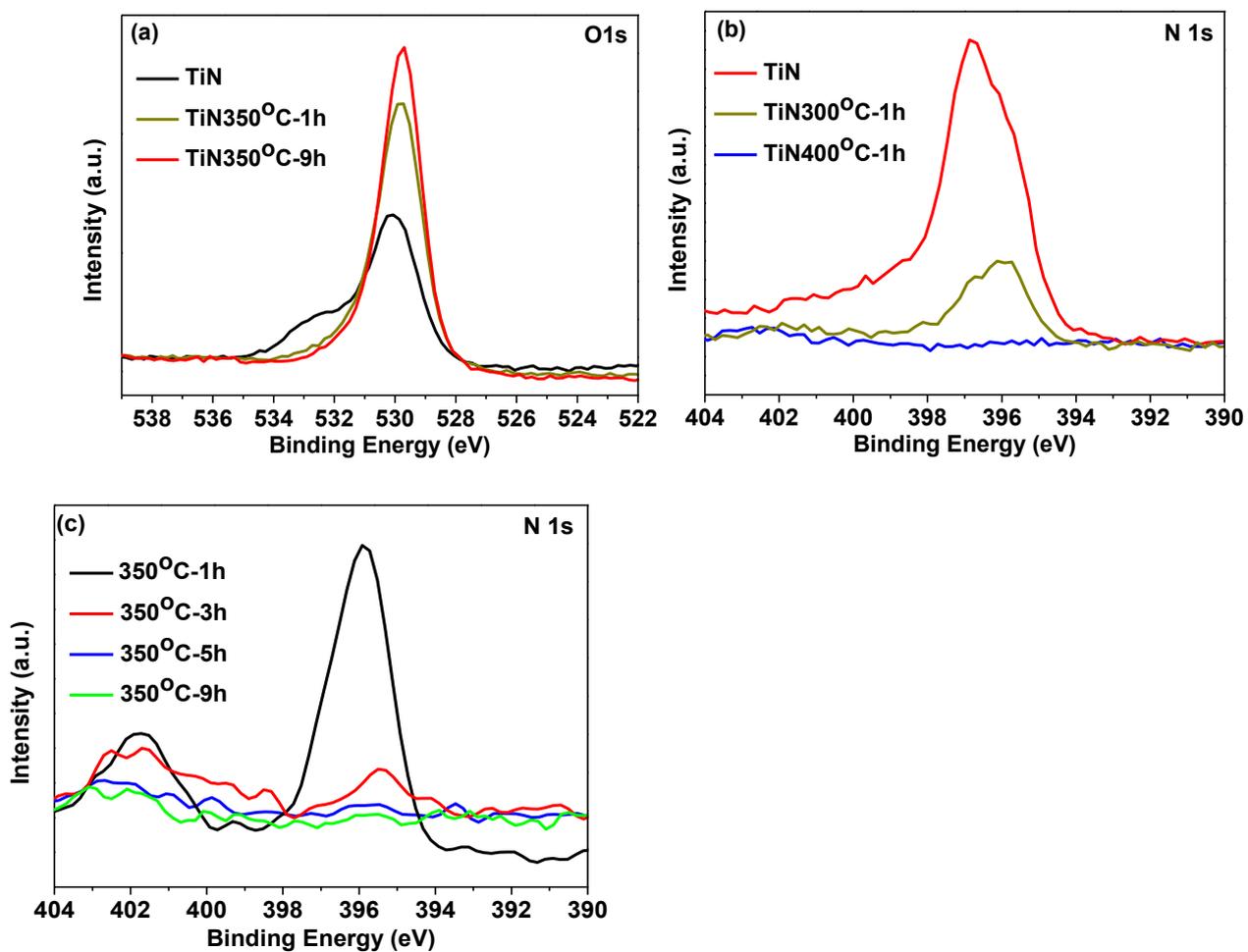

Fig. S5. High resolution XPS O1s peaks (a) and N1s peaks (b/c) for TiN nanopowders and oxidized TiN at 300°C to 400°C for 1h to 9h, respectively.

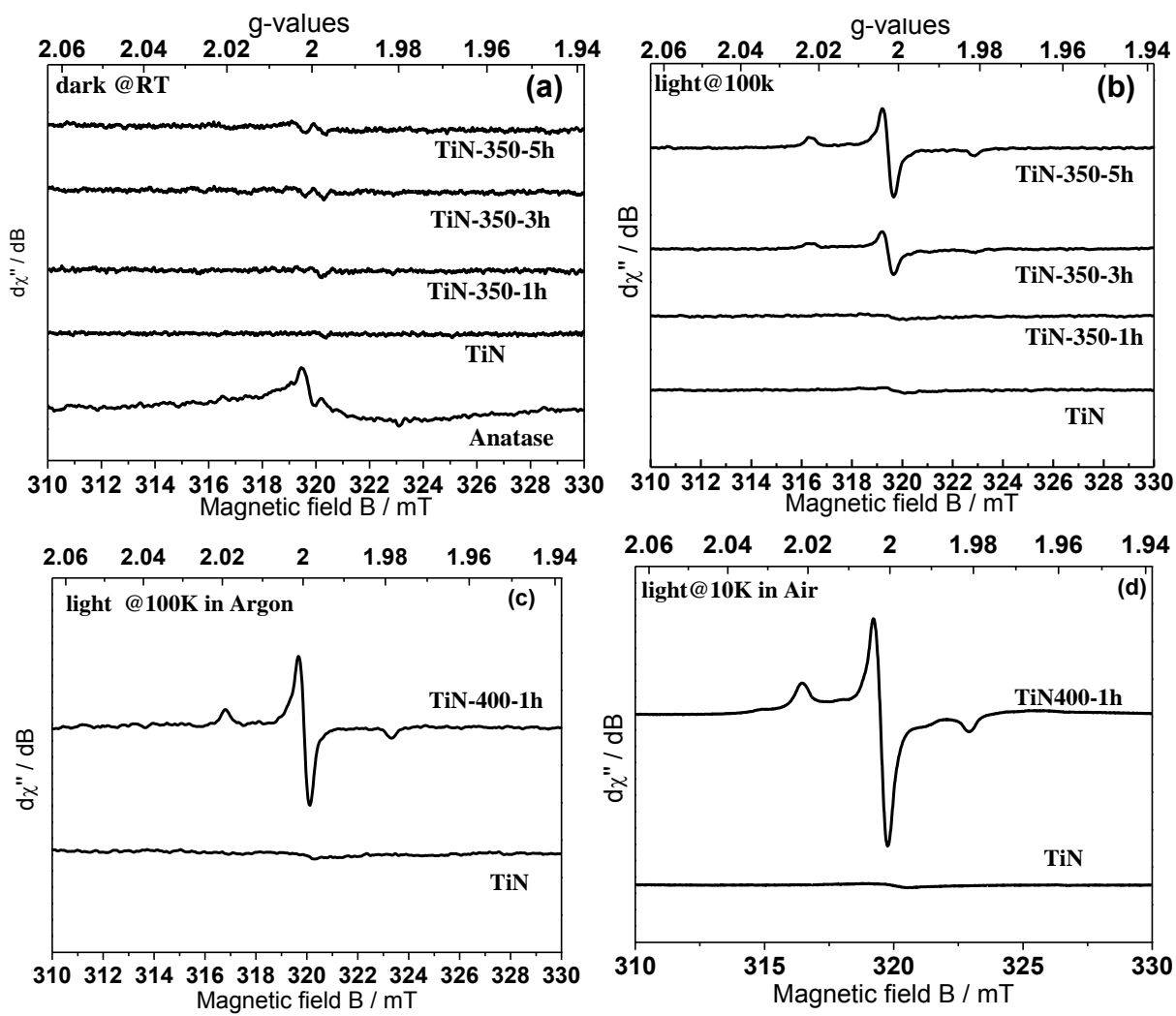

Fig.S6. EPR spectra of (a-b) titanium nitride and oxidized titanium nitride at 350 °C for 1 h, 3 h and 5 h. The spectra were taken at RT, and at 100 K under illumination in an air atmosphere, respectively. EPR spectra of oxidized TiN (400 °C, 1h) and TiN (c) at 100 K under illumination in argon atmosphere and (d) at 10 K under illumination in air atmosphere.